\documentclass[10pt,a4paper,showpacs,english,superscriptaddress,nofootinbib,twocolumn, reprint]{revtex4-1}
\pdfoutput=1
\usepackage{color}
\usepackage{amssymb}
\usepackage{amsmath}
\usepackage{graphicx}
\usepackage{babel}
\usepackage[T1]{fontenc}
\usepackage{hyperref}
\usepackage{amsfonts}
\usepackage{caption}

\newcommand{\beq}{\begin{equation}}

\newcommand{\eeq}{\end{equation}}
\newcommand{\beqa}{\begin{eqnarray}}
\newcommand{\eeqa}{\end{eqnarray}}

\graphicspath{{images/}{../images/}}
\def\p{\mbox{\boldmath$\displaystyle\mathbf{p}$}}
\def\0{\mbox{\boldmath$\displaystyle\mathbf{0}$}}

\def\to{\rightarrow}
\def\gev{\hbox{GeV}}

\newcommand{\mawave}{{\mathcal G}} % Majorana wave operator
\def\eslash{\not\!\! E_T}

\begin{document}
\title{Searching for Elko dark matter spinors at the CERN LHC}

\author{Alexandre Alves}
\email{aalves@unifesp.br}
\affiliation{Departamento de Ci\^encias Exatas e da Terra,Universidade Federal de S\~ao Paulo\\ Diadema-SP-Brazil}

\author{F. de Campos}
\email{fernando.carvalho@ict.unesp.br}
\affiliation{Departamento de Engenharia Ambiental, ICT-Universidade Estadual Paulista,\\
 S\~ao Jos\'e dos Campos-SP-Brazil}

\author{M. Dias}
\email{marco.dias@unifesp.br}
\affiliation{Departamento de Ci\^encias Exatas e da Terra,Universidade Federal de S\~ao Paulo\\ Diadema-SP-Brazil}

\author{J. M. Hoff da Silva} 
\email{hoff@feg.unesp.br; hoff@ift.unesp.br}
\affiliation{Departamento de F\'isica e Qu\'imica, Universidade Estadual Paulista,\\
 Guaratinguet\'a-SP-Brazil}


\begin{abstract}
The aim of this work is to explore the possibility to discover a fermionic field with mass dimension one, the Elko field, in the Large Hadron Collider (LHC).  Due to its mass dimension, an Elko can only
interact either with Standard Model (SM) spinors and gauge fields at 1-loop order or at tree level through a quartic interaction with the Higgs field. In this Higgs portal scenario, the Elko is a viable candidate to a dark matter constituent which has been shown to be compatible with relic abundance measurements from WMAP and 
direct dark matter--nucleon searches. We propose a search strategy for this dark matter candidate in the channel $pp\rightarrow \ell^+\ell^-+\eslash$  
at the $\sqrt{s}=14$  TeV LHC. 
We show the LHC potential to discover the Elko considering a triple Higgs-Elko coupling as small as $\sim 0.5$ after 1 ab$^{-1}$ of integrated luminosity. Some phenomenological consequences of this new particle and its collider signatures are also discussed.  
\end{abstract}

\maketitle
\noindent

\section{Introduction}

The so-called Elko spinor fields are a set of four spinors, whose main characteristic is to be eingenspinors of 
the charge conjugation operator. This construction lead these spinors to a very exhaustive property: they are 
blind with respect to electromagnetic interactions. Moreover, the (necessary) quantum field associated to 
Elko spinors presents the peculiarity of having mass dimension one, severely reducing the perturbatively
renormalizable possible couplings of the quantum Elko field \cite{elko}. These two characteristics underlining the 
formal Elko structure are responsible for making the spinor deserve the epithet of {\it a first principle dark 
matter candidate}.  
A complementary analysis involving relic density and gravitational collapse of a primordial Elko could give together 
the Elko mass parameter of the order of a few MeV \cite{elko2} for the Elko field viability as a physical dark 
matter candidate.      

With the CERN LHC in operation and and its success in the recent discovery of the Higgs field, it is natural to ask 
whether the features of the Elko field can be probed on such accelerator. In this work we investigate the possibility for Elko production at the CERN LHC, at 14 TeV, through the process $pp\rightarrow Z^*\rightarrow Z+
\hbox{elkos}\to \ell^+\ell^-+\eslash$, where $\ell=e,\mu$.

We are dealing with processes not shared by other fields in the Standard Model: the specific vertex containing two 
scalar and two fermionic fields has mass dimension five in the Standard Model, requiring a negative mass 
dimension coupling constant, rendering the interaction non renormalizable perturbatively. In other words, due to the 
unusual Elko mass dimension, it is possible to investigate processes which are {\it ab initio} beyond the standard 
model. The difficulties in dealing with such processes rest on the fact that they are mediated by at least one loop 
in neutral scalar and, besides, we have faced a huge background.

In Section II we shall briefly resume the Elko construction and some of its properties. In Section III, we make the 
phenomenological analysis of the two aforementioned channels, comparing them, and presenting some estimates for the 
relevant coupling constants. In Section IV we present our conclusions.

\section{Elko spinor field}

The charge conjugation operator, in the Weyl representation, is given by
\beq
{C} =
\left(
\begin{array}{cc}
\mathbb{O} & i\,\Theta \\
-i\,\Theta & \mathbb{O}
\end{array}
\right) {K}\,,\label{cc}
\eeq
where $K$ is responsible for complex conjugate the spinor it is acting on and $\Theta=-i\sigma_2$. The very equation 
defining the Elko spinors is given by 
\beq
{C} \lambda(\p) = \pm  \lambda(\p). \label{eq:neutral}
\eeq It is possible to show that there are four eigenspinors, two corresponding to the $+$ plus --- the self 
conjugated spinors $\lambda^S$ --- and two related to the $-$ sign, the anti-self conjugated spinors. The functional 
form for these spinors may be found explicitly in, for instance, \cite{elko,elko2}. Here, we shall focusing in the 
main properties concerning our analysis. 
We shall pinpoint that the explicit construction of the Elko spinors defines it as a member of all 
the $r\oplus l$ Weyl representation space, i. e., the spinor carries both 
helicities. 

It is possible to find the correct dual for the Elko spinor by means of a quite precise criteria, as follows. Let us 
demand that the product $[\lambda_\alpha]^\dagger \eta \lambda_\beta$ be invariant under arbitrary Lorentz 
transformations, where the index labels one of the four Elko. This requirement amounts out as the constraints 
$[J_i,\eta]=0=\{K_i,\eta\}$, being $J$ and $K$ the Lorentz transformation generators of rotations and boosts, 
respectively. It can be readily verified that the unique consistent solution is given by \cite{elksplb} 
$\eta=\pm i\gamma^0$ and the dual representation so that the norm is well-defined (leading to a positive definite 
hamiltonian) is given by
\beq
 \stackrel{\neg}\lambda^{S/A}_{\{\mp,\pm\}}(\p)
:= \pm \,i\,\left[\lambda^{S/A}_{\{\pm,\mp\}}(\p)\right]^\dagger\gamma^0\,.
\eeq
With the aid of the above equations, it is possible to set down the orthonormality relations:
\begin{subequations}
\begin{align}
& \stackrel{\neg}\lambda^S_{\alpha}(\p)\;
\lambda^I_{\alpha^\prime}(\p) = +\; 2 m\; \delta_{\alpha\alpha^\prime}\;\delta_{SI}\,,
\label{zd1}\\
& \stackrel{\neg}\lambda^A_{\alpha}(\p)\;
\lambda^I_{\alpha^\prime}(\p) = -\; 2 m \;\delta_{\alpha\alpha^\prime}\;\delta_{AI}\,,
\label{z5}
\end{align}
\end{subequations}
where $I\in\{S,A\}$ and the completeness relation 
\beq
\frac{1}{2 m}\sum_\alpha 
 \left[\lambda^S_{\alpha}(\p) \stackrel{\neg}\lambda^S_{\alpha}(\p) 
      - \lambda^A_{\alpha} (\p)\stackrel{\neg}\lambda^A_{\alpha}(\p)\right]
 = \mathbb{I}\,, \label{z1}
\eeq with $\alpha= \{+,-\}, \{-,+\}$.

It is important to emphasize the emergence of unusual {\it spin sums}, given by  
\begin{subequations}
\beqa
&&  \sum_{\alpha} \lambda^S_\alpha(\p)\stackrel{\neg}\lambda_\alpha^S(\p) 
= +\, m\big[ \mathbb{I}+\mawave(\p)\big]\,, \label{spinsumS}\\
&&  \sum_{\alpha} \lambda^A_\alpha(\p)\stackrel{\neg}\lambda_\alpha^A(\p) 
=  - m\,\big[\mathbb{I}-{\mawave(\p)}\big]\,; \label{spinsumA}
\eeqa
\end{subequations} where one can write down the explicit form for $\mawave(\p)$ as
\begin{eqnarray*}
\mawave(\p)=\left(\begin{array}{cccc} 0&0&0&-ie^{-i\phi}\\0&0&ie^{i\phi}&0\\0&-ie^{-i\phi}&0&0\\ie^{i\phi}&0&0&0 
\end{array}\right).
\end{eqnarray*} Notice that the spin sums are modified by a rather non trivial term whose argument is given by the 
momentum. These spin sums do violate (in a rather subtle way) the full Lorentz invariance, making Elko fields 
invariant under a subgroup of the Lorentz group (see reference \cite{ult} for an up to date account on the 
formalism). We remark that in the subgroup case in question, rotations and boosts still present as generators.  

It is easy to show that the Elko spinor satisfy a Dirac-like equation which is only an algebraic identity (nothing 
to do with the field dynamics) given by
\[\mathcal{D}\lambda_\beta^{S/A}(\p)=\left(\gamma_\mu p^\mu \delta_\alpha^\beta \pm im\mathbb{I}
\varepsilon_\alpha^\beta\right)\lambda_\beta^{S/A}(\p)=0,\]
where $\delta^\alpha_\beta$ is the usual Kronecker symbol and the antisymmetric symbol $\varepsilon$ is defined as 
$\varepsilon^{\{-,+\}}_{\{+,-\}}:=-1$. The importance of such a relation shall not be underestimated. In fact, the 
very existence of the operator $\mathcal{D}$ acting on the spinor space gives information about the physical content 
encoded on the Elko spinor: the covariance condition arising from the $\mathcal{D}$ operator is the same of the 
Dirac one and, therefore, the corresponding transformation on $\lambda^{S/A}$ is not unitary. As a result, 
$\lambda^{S/A}$ cannot be associated to a quantum state in any sense and (second) quantization is necessary.   

The full consistent quantum field associated with the Elko can be written as \cite{elko,elko2}
\begin{eqnarray}
%\beq
 \eta(x)&=&\int \frac{d^3 p}{(2\pi)^3}\frac{1}{\sqrt{2\, m\, E(\p)}} 
\sum_{\beta}   \Big[c_\beta(\p) \lambda^S_\beta(\p)  \mathrm{e}^{-ip_\mu x^\mu}\nonumber\\
&+& c_\beta^\dagger(\p)  
\lambda^A_\beta(\p) \mathrm{e}^{+ip_\mu x^\mu}\Big]\,.\nonumber
%\eeq
\end{eqnarray}
Analogously, its dual (${\stackrel{\neg}\eta}(x)$) is obtained by replacing $\lambda$ for 
${\stackrel{\neg}\lambda}$, $c$ for $c^\dagger$ and $ip_\mu x^\mu\leftrightarrow-ip_\mu x^\mu$. 
The anticommutators for the creation and destruction operators,  $c^\dagger_\beta(\p)$ and $c_\beta(\p)$, are:
\begin{align}
& \big\{c_\beta(\p),\; c^\dagger_{\beta^\prime}(\p^\prime)\big\}
=  \left(2\pi\right)^3  \delta^3\left(\p-\p^\prime\right)
\delta_{\beta\beta^\prime}\,,\label{eq:anticomm} \\
&\big\{c^\dagger_\beta(\p),\; c^\dagger_{\beta^\prime}(\p^\prime)\big\}
=\big\{c_\beta(\p),\; c_{\beta^\prime}(\p^\prime)\big\}=0\,.
\end{align}

In order to unveil the dynamics associated to the quantum field, it is necessary a bottom-up approach. The best 
procedure is to calculate the Feynman-Dyson propagator inferring, then, the corresponding lagrangian. After a slightly 
modified textbook calculation, one arrives at
\beq
\mathcal{S} (x-x^\prime) = \int\frac{d^4 p}{(2\pi)^4}\;
\mathrm{e}^{i p_\mu(x^\mu-x^{\prime\mu})}\;
\frac{\mathbb{I}}{p_\mu p^\mu - m^2 + i\epsilon}\, ,
\eeq which is nothing but the Klein-Gordon propagator. Hence, the Elko spinor field has mass dimension one and 
satisfy the Klein-Gordon equation,
\[(p_\mu p^\mu-m^2)\lambda^{S/A}(\p)=0.\] As consequence, to only perturbatively renormalizable terms in the 
Lagrangian density are the mass term and an interaction of a scalar field
\begin{eqnarray*}
  \mathcal{L} = \partial^\mu {\stackrel{\neg}\eta}(x) \,\partial_\mu\eta(x) - m_\varepsilon^2\stackrel{\neg}\eta(x)\,
  \eta(x) +\lambda_E \eta(x)\stackrel{\neg}\eta(x)\phi(x)^2,
\end{eqnarray*}
where $\lambda_E$ is the coupling constant. Thus, the only possible interaction with vector bosons is via production 
of two Higgs mediated by a loop of these particles, with a subsequent generation of a $ \eta\stackrel{\neg}\eta$ 
pair, in the form of an effective vertex.

Nevertheless, we can also shift the Higgs by a non zero vacuum expectation value (VEV), obtaining a triple Higgs(H)-Elko(E)-Elko(E), $HEE$, vertex
\begin{equation}\label{marado1}
\mathcal{L}_{int}=\alpha_E \eta(x)\stackrel{\neg}\eta(x)\phi(x),
\end{equation}
with $\left[\alpha_E\right]=\left[mass\right],$ also renormalizable. This coupling constant naturally arises in a 
theory where Elkos also get their masses from the electroweak symmetry breaking mechanism, and its relation with 
$\lambda_E$ and the Higgs VEV is
\[\alpha_E=\frac{v}{\sqrt{2}}\lambda_E.\]
 As we are going to show, Elkos that couple to Higgs bosons 
 according to Eq.~\ref{marado1} have a relevant production cross section at the 14 TeV LHC, unlike those that couples to Higgs bosons through quartic interactions only.
 
 \subsection{Computation of the 1-loop amplitudes}
 
We compute the 1-loop amplitudes contributing to this production mode, that will be extended for all analysis containing a $\lambda_E$ coupling. Afterwards we propose a search strategy for Elkos at the 14 TeV LHC.

\begin{center}
\begin{figure}[h!]
	\includegraphics[scale=0.27]{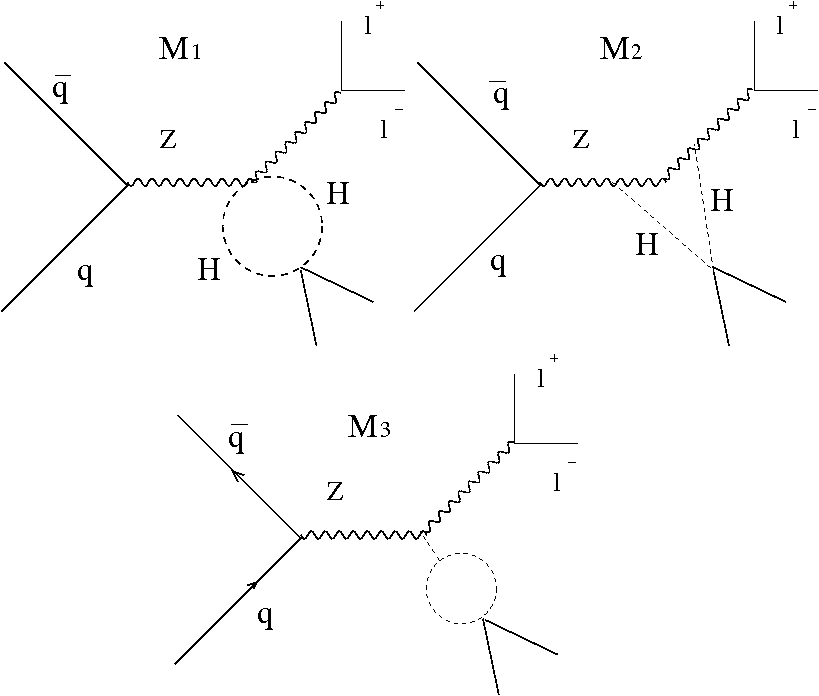}
	\caption{Feynman diagram for the production of a pair of
Elko spinors associated with a SM gauge boson $Z_0$.}
	\label{figfeyn}
	\end{figure}
\end{center}

The Feynman diagrams contributing to the process $pp\to Z+E\overline{E}$ is shown in Figure~\ref{figfeyn}.

The effective coupling between two $Z$ bosons and two Elko fields, which gives rise to the amplitude $M_1$ in 
Figure~\ref{figfeyn}, in the on shell renormalization scheme is given by
($\tilde{V}_{ZZ\to \lambda\stackrel{\neg}{\lambda}}=V^{eff}_{ZZ\to \lambda\stackrel{\neg}{\lambda}}(p_5,p_6)
 -V^{eff}_{ZZ\to \lambda\stackrel{\neg}{\lambda}}(0)$), due a Higgs loop, using a 
 cutoff scale,
\begin{eqnarray}\label{byfrost1}\nonumber 
\tilde{V}^{eff}\!\!&=&\!\!\left\{\begin{array}{cc}\!\!\!\!\!\!\!\!\!\!\!\!\! g_1\frac{\lambda_E}{16\pi^2}\left(2+
\sqrt{1-\gamma}\; \ln\left|\frac{1-\sqrt{1-\gamma}}{1+\sqrt{1-\gamma}}\right|\right)+r\!\!\!\!\!\! & \!\!\!\!\!\!\!
\!\!\!\!\!\!\!\!\!\!\!\!, \,  \gamma \leq 1\\
g_1\frac{\lambda_E}{16\pi^2}\left(2-2\sqrt{\gamma-1}\;\tan^{-1}(\sqrt{\gamma-1})\right)+r &\!\!\!\! , \,  \gamma >1
\end{array}\right.\\
\end{eqnarray}
where $g_1$ is the coupling between two Higgs and two $Z$ bosons, $\gamma=\frac{4m_H^2}{|p_5+p_6|^2}$ and $p_{5,6}$ 
are the final state Elkos four-momentum. The $r$ function is
\begin{eqnarray*}r&=&g_2\frac{\lambda_E}{16\pi^2}\int_0^1 dx \frac{\tan^{-1}\left(\frac{(q+|p^5+p_6|)x+|p_5+p_6|}
{\Delta}\right)}{|p_5+p_6| \Delta}\\
&-&\frac{\tan^{-1}\left(\frac{(q+|p_5+p_6|)x}{\Delta}\right)}{|p_5+p_6| \Delta}, \;\;\; \textnormal{where}\\
\Delta&=&\sqrt{-m^2_H(x-1)+m^2_Z x-(q+\frac{|p^5+p_6|}{2})^2x},
\end{eqnarray*}
depending on the $Z$ mass,  $m_Z$, the coupling between one Higgs and two $Z$ bosons, $g_2$, and the quartic $HHEE$ 
coupling $\lambda_E$. The external momentum of a vector boson entering in the effective vertex is denoted by $q$.

The second amplitude contributing to the effective vertex, $M_2$, is a tadpole diagram involving one Higgs and two 
Elkos, and is given by
\begin{eqnarray}\label{byfrost5}\nonumber
\tilde{V}^{eff}_{H\to \lambda\stackrel{\neg}{\lambda}}&=&\frac{\lambda_E}{16\pi^2}\left(2-\frac{2m_H\, tan^{-1}
\left(\frac{|p_5+p_6|}{m_H}\right)}{|p_5+p_6|}\right.\\
&+&\left.\; \ln\left|\frac{m^2_H}{m^2_H+|p_5+p_6|^2}\right|\right).
\end{eqnarray}

We also will need the spin sum of dimension-one Elko spinors  to compute the square amplitude to the partonic process 
$q\bar{q}\to Z$ and two Elkos depicted in Figure~\ref{figfeyn}. To perform it we use the fact that  ~\cite{elko}
\begin{eqnarray}\label{byfrost6}\nonumber
\lambda^{S/A}_{\left\{\pm, \mp\right\}}(-\vec{p})&=&\pm i \lambda^{A/S}_{\left\{\mp, \pm\right\}}(\vec{p}),
\end{eqnarray}
since $\lambda^{S/A}_{\left\{\pm, \mp\right\}}(\vec{0})=\pm i \lambda^{A/S}_{\left\{\mp, \pm\right\}}(\vec{0})$ when $\phi\to \phi+\pi$ and $\theta\to \pi-\theta$ and  the  Lorentz boost to $\vec{p}$ depends only on $p=|\vec{p}|$. Consequently
\begin{eqnarray}\label{bred6}\nonumber
\stackrel{\neg}{\lambda}^{S/A}_{\left\{\pm, \mp\right\}}(-\vec{p})&=& \pm i \stackrel{\neg}{\lambda}^{A/S}_{\left\{\mp, \pm\right\}}(\vec{p}),
\end{eqnarray}
or, 
\begin{eqnarray}\label{bred6}\nonumber
\stackrel{\neg}{\lambda}^{S/A}_{\alpha}(-\vec{p})&=&- i\varepsilon^\beta_\alpha \stackrel{\neg}{\lambda}^{A/S}_{\beta}(\vec{p}).
\end{eqnarray}
Let us fix the spinor indices $a,b=1,2,3,4$ to write the amplitude of the simplest scattering process -- the two Higgs annihilation generating two Elkos -- as
\begin{eqnarray}\nonumber
\mathcal{M}&=&\frac{\lambda_E}{m_\varepsilon}\lambda^{I,a}_\alpha(\vec{p})\stackrel{\neg}{\lambda}_{\alpha^\prime}^{J,b}(\vec{-p})
\delta^{ab},
\end{eqnarray}
so that
\begin{eqnarray}\nonumber
|\mathcal{M}|^2&=&\frac{\lambda_E^2}{m^2_\varepsilon}\lambda^{I,a}_\alpha (\vec{p}) {\lambda^{I,c}_\alpha}^\dagger (\vec{p})
\stackrel{\neg}{\lambda}^{J,b}_{\alpha^\prime}(-\vec{p}){\stackrel{\neg}{\lambda}^{J,d}_{\alpha^\prime}}^\dagger(-\vec{p})\delta^{ab}
\delta^{cd}.
\end{eqnarray}
The spin and Elko type averages are
\begin{eqnarray}\label{byfrost2}\nonumber
\frac{1}{16}\displaystyle\sum_{\alpha,\alpha^\prime}\displaystyle\sum_{I,J}|\mathcal{M}|^2&=& \frac{\lambda_E}{16m_
\varepsilon^2}\displaystyle\sum_{\alpha, \alpha^\prime}(\lambda^A_\alpha{\lambda_\alpha^A}^\dagger+\lambda^S_
\alpha{\lambda_\alpha^S}^\dagger)_{ac}\\\nonumber
&\times&\displaystyle\sum_{I}(\stackrel{\neg}{\lambda}_{\alpha^\prime}^I{\stackrel{\neg}{\lambda}_{\alpha^\prime}^I}
^\dagger)_{bd}\delta^{ab}\delta^{cd}.
\end{eqnarray} The sum in the first parenthesis of Equation (\ref{byfrost2}) is given by $2(E\mathbb{I} -P
\mawave)_{ac}$ (see Equations (B.18) and (B.19) of \cite{elko}), while in the second line we have
\[\displaystyle\sum_{\alpha^\prime}\stackrel{\neg}{\lambda}_{\alpha^\prime}^I{\stackrel{\neg}{\lambda}_{\alpha^
\prime}^I}^\dagger=\displaystyle\sum_{\alpha}{\lambda_{\alpha^\prime}^I}^\dagger\lambda^I_{\alpha^\prime}.\] Thus
\begin{eqnarray}\label{byfrost3}\nonumber
\frac{1}{16}\displaystyle\sum_{\alpha,\alpha^\prime}\displaystyle\sum_{I,J}|\mathcal{M}|^2&=&\frac{\lambda^2_E}
{8m^2_\varepsilon}\left(E\delta_{bd}-p\mawave_{bd}\right)\displaystyle\sum_{I}\displaystyle\sum_{\alpha}
{\lambda^{I,b}_\alpha}^\dagger\lambda^{I,d}_\alpha.\\
\end{eqnarray}
Now making use of Equations (B.24) and (B.25) of reference \cite{elko} 
\[\displaystyle\sum_{\alpha}{\lambda_\alpha^I}^\dagger\lambda_\alpha^I=2(E-p)+2(E+p)=4E,\]
we can write 
\begin{eqnarray}\label{byfrost4}\nonumber
\frac{1}{16}\displaystyle\sum_{\alpha,\alpha^\prime}\displaystyle\sum_{I,J}|\mathcal{M}|^2&=&\frac{\lambda^2_E}
{16m^2_\varepsilon}\left[16E^2-2p\displaystyle\sum_{I}\displaystyle\sum_\alpha \left({\lambda^{I,b}_\alpha}^\dagger 
\mawave_{bd}\lambda^{I,d}_\alpha\right)\right]\\
&=&\frac{\lambda^2_E}{16 m^2_\varepsilon}\left[16E^2-2p\, tr\left(2E\mawave +2p\mathbb{I}\right)\right]\nonumber\\ 
&=&\frac{\lambda^2_E}{m^2_\varepsilon}(E^2-p^2)\nonumber\\
&=&\lambda_E^2,
\end{eqnarray} where the ciclicity property of the trace on the first line was used. 

Combining the Equations (\ref{byfrost1}), (\ref{byfrost5}) and (\ref{byfrost4}) one can construct an effective 
vertex relating two vector bosons and two Elko fields to obtain relevant signal amplitudes.

\section{Searching for Elkos at the LHC}

\subsection{Quartic coupling scenario}
Consider only the $\lambda_E$ type coupling. As a consequence of its sole tree-level coupling being a quartic coupling to Higgs bosons pairs, Elkos can only be 
pair produced at colliders in this scenario. The associated $gg\to H^*\to H+E\overline{E}$ production is the most 
straightforward tree-level mechanism to produce Elkos but with a low cross section, considering the Higgs boson decay. 

However at 1-loop level several possibilities are opened. For example, a Higgs boson might decay to Elkos pairs leading to monojet and monophoton signatures. The quartic $VVHH$, $V=W,Z$, couplings might give rise to weak boson fusion (WBF) $jj+E\overline{E}$ and associated $V=Z,W$ boson-Elkos $V+E\overline{E}$ events by their turn. We are going to discuss the difficulties associated to these channels now.

\vskip0.5cm
\underline{Associated $Z$-Elkos channel}
\vskip0.5cm

Let us analyze the associated production of a $Z$ boson and two Elkos in the channel $\ell^\pm\ell^\mp+\eslash$, $\ell=e,\mu$. The $W$-Elkos is a more difficult channel because it is much harder to identify the signal events based solely on missing energy distributions. 

We simulate the signal events with \verb+Madgraph5+ \cite{madgraph} modifying the SM $ZZHH$ and $ZZH$ vertices to take into account the effective vertices of the Elko case. We stress the fact that the vertex between two Higgs 
particles and $ \lambda^{I,a}_\alpha$ and $\stackrel{\neg}{\lambda}_{\alpha^\prime}^{J,b}$, namely
\[\lambda_E \delta^{ab},\]
was chosen in order to maintain this coupling renormalizable. Also we set the Elko mass to $m_\varepsilon=0.01 GeV$, in consonance with the Elko mass range estimated in dark matter direct and indirect searches~\cite{elko}. 

The total cross section for Elko production in association to a $Z$ boson is around $10^{-5} \, \hbox{fb}$ at the 8 
TeV LHC, which is far beyond the LHC reach for the current integrated luminosity. Thus we proceed to the prospects 
for Elko discovery at the 14 TeV LHC. The factorization scale is set to be $\sqrt{\hat{s}}$, $\hat{s}$ being the 
parton level center of mass energy. We do not expect any large deviations of our partonic level estimates by 
including detector effects and showering, given the optimal coverage and detection efficiency of the LHC detectors 
for events with hard electrons and muons, and large missing transverse energy. Yet, a full simulation should be 
performed to properly evaluate those effects. 

In the process of Fig.(\ref{figfeyn}) the final state of interest consists of two opposite-sign electrons or muons 
and missing energy associated to the Elkos production. The main background contributions come from~\cite{Trefzger:1999bna}:
\begin{enumerate}\label{6}
\item {(1)} $ZZ,Z\gamma\to \ell^+ \ell^-+\nu_\ell \bar{\nu}_\ell$;
\item {(2)} $W^+ W^-\to \ell^+\ell^- +\nu_\ell \bar{\nu}_l$;
%\item $ Z \, \gamma \to \nu \bar{\nu}_l l^+ l^-$
\item {(3)} $W^\pm \, Z \to \ell^\pm \ell^\mp \ell^\pm +\nu_\ell$ with one missing charged lepton;
\item {(4)} $t\bar{t}\to W^+ W^-b \bar{b}\to \ell^+\ell^- b \bar{b}+\nu_\ell \bar{\nu}_\ell$;
\item {(5)} $W^\pm j \to \ell^+\ell^- + \nu_\ell$, with the jet misidentified as a charged lepton in the detector.
\end{enumerate} We also have generated all these samples using \verb+Madgraph5+. 

The reducible backgrounds $W^+ \, Z \to l^+ l^- l^+ \nu_l$ and $ W^- \, Z \to l^- l^+ l^- \bar{\nu}_l $ are suppressed imposing only two opposite sign leptons on the final state since a third charged lepton is rarely outside the fiducial region of the detectors. The acceptance cuts for charged leptons are given by
\begin{equation}
p_{T_\ell} > 10\; \hbox{GeV} \;\; ,\;\; |\eta_\ell| < 2.5
\label{accept}
\end{equation}

Fake leptons are another source of potential background events which arise from a jet enriched environment. In order 
to estimate the probability of a jet to be misidentified as a lepton we simulated a sample using \verb+Pythia+  
\cite{pythia} for jet showering and clustering, and  \verb+PGS+ for detector simulation, for different 
signal leptons on the final state. We have found a probability of $10^{-4}$ for a jet to be misidentified as an 
isolated lepton, which is consistent to the presented experimental studies~\cite{atlas,atlas2}. Using this result, 
the background $W^\pm+j$ can be eliminated.
 
In order to increase the signal to background ratio, we demand the events to satisfy the following cut 
\begin{equation} \label{cuts2}
\displaystyle\sum_{visible}|\vec{p}_T| < 120\gev ,
\end{equation}
where $\displaystyle\sum_{visible}|\vec{p}_T|$ is the scalar sum of the transverse momentum vector of all visible objects, in the case of our signal, the charged leptons. Unfortunately, even after this cut a signal to background ratio relevant for discovery at the LHC seems difficult. The cross sections are described on Table (\ref{tab01}).
\begin{center}
\begingroup
\squeezetable
\begin{table}
\begin{tabular}{|c|c|}\hline
Process & $\sigma (pb)$ \\\hline
$pp\to \ell^+\ell^-+\eslash$ & \\\hline
Signal & $4.69\times 10^{-4}$ \\\hline
Background & $1.969$\\\hline\hline
$pp\to jj+\eslash$(WBF) & \\\hline
Signal & $0.001119$\\\hline
Background & $0.75$ \\\hline\hline
$pp\to j+\eslash$ & \\\hline
Signal & $0.02531\times 10^{-5}$\\\hline
Background & $2820$ \\\hline\hline
\end{tabular}
\caption{Cross-sections for signal and backgrounds for three of the most promising channels for Elko discovery at the 14 TeV LHC. Cuts applied to reduce backgrounds are described in the text.}
	\label{tab01}
\end{table}
\endgroup
\end{center}

\vskip0.5cm
\underline{Elkos in Weak Boson Fusion}
\vskip0.5cm

Concerning the Elko plus two jets signal, the WBF channel, we used the \verb+PGS+ and \verb+Pythia+ on the samples to make our analysis more reliable. We have applied the following cuts on both signal and background:
\begin{eqnarray*}
|\eta_j|&<& 2.5, p_{T_j}> 30\, GeV \, \textnormal{ and } M_{jj}> 800\, GeV,
\end{eqnarray*}
where $M_{jj}$ in the jets invariant mass.

Table (\ref{tab01}) summarizes our results. The background quoted in the table is the irreducible one $pp\to jj\nu_\ell\overline{\nu}_\ell$.
Considering the large background, again we are not able to obtain a relevant signal to background significance ratio for this channel at the LHC.

\vskip0.5cm
\underline{Elkos in Gluon Fusion}
\vskip0.5cm

\begin{figure}
\begin{center}
\includegraphics[scale=.4]{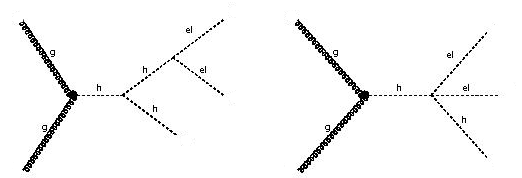}
\end{center}
\caption{Process using the effective vertex for Higgs production.}\label{tho2}
\end{figure}

Using a modified version of \verb+heft+ model on \verb+MadGraph+ we also obtained the cross section for the process on Figure \ref{tho2}, where an off-shell Higgs boson is produced in gluon fusion and decays to Elkos and an additional Higgs boson,
$\sigma=7.183\times 10^{-5}pb$, for $m_\varepsilon=10$ MeV and coupling 
constant between two Higgs and two Elkos set to one. This is a too low cross-section to proceed with a detailed analysis.

\vskip0.5cm
\underline{Monojet channel}
\vskip0.5cm

Now we look for the possibility to observe Elkos in one jet plus missing energy channel. This monojet channel has been intensely studied as a promising way for dark matter detection at hadron colliders~\cite{monoj}. On Table (\ref{tab01}) we quote the cross-sections for signal and background in this case. As we can see, the prospects to get a reasonable signal to background ratio for discovering Elko through this signature are hopeless.

After analyzing the most relevant channels to Elko discovery through quartic interactions with discouraging results, let us now investigate the scenario where the Elko possess triple couplings to Higgs bosons.

\subsection{Triple coupling scenario}

The situation is very different for the coupling described by Eq. (\ref{marado1}). Again we simulate the $pp \to l^+ l^- +\eslash$. The signal cross section due the $HEE$ coupling is $0.103$ pb, after acceptance cuts of Eq.~(\ref{accept}), this time for 10 MeV Elkos and $\alpha_E=1$. This increase in the production cross section is consequence of the tree-level coupling involved in the Elkos production. The Higgs branching ratio to Elkos is around 6\% which is still comfortably allowed by the LHC data~\cite{invhiggs}.

After demanding the acceptance cuts of Eq.~(\ref{accept}) and the cut of Eq.~(\ref{cuts2}), we have implemented the following additional cuts (\ref{cutsmar}),
\begin{eqnarray}
80\gev &<& M_{\ell\ell} <  120\gev ,\\
\eslash &>& 50\gev .
\label{cutsmar}
\end{eqnarray}
The cut on the leptons invariant mass, $M_{\ell\ell}$, eliminates the events where the lepton pair is not produced by an on-shell $Z$ boson including the subdominant irreducible backgrounds $Z\gamma$ and $W^+W^-$. By the 
way, this is the reason why we chose only the associated production of a $Z$ boson and Elkos whereas it would be possible to include the $W$ plus Elkos production as well. As a leptonic $W$ cannot be reconstructed and the Elko 
events do not present a large amount of missing energy when compared to the resonant $W$ production, the SM background $pp\to W^\pm\to \ell^\pm+\eslash$ would be overwhelming. The missing energy cut, by its turn, is essential for trigger purposes and helps to increase the signal to background ratio.

After imposing this cut we obtained the results in Table (\ref{tab1}). The signal acceptance is 0.63 after applying all cuts.
\begin{center}
\begingroup
\squeezetable
\begin{table}[h!]
	\begin{tabular}{|c|c|}\hline
Process&  $\sigma$(fb) after all cuts \\\hline
Signal & $6.8$\\\hline
Background & $133.6$\\\hline
\end{tabular}
\caption{Cross-sections for signal and background in $pp\to \ell^+\ell^- +\eslash$, $\ell=e,\mu$, after all cuts in the triple coupling scenario.}
	\label{tab1}
\end{table}
\endgroup
\end{center}

Using $\sqrt{2((S+B)ln\left(1+\frac{S}{B}\right)-S)}$, $S$ and $B$ the number of signal and backgrounds events, respectively,  as the test 
statistic and considering that the number of events is quadratic in the coupling constant, we obtain the statistical significance in terms of  the coupling constant and the integrated luminosity, $L_{int}$. 

We show in Figure~\ref{lumilam} the required luminosity to a $3,5$ and $10\sigma$ signal as a function of the $\alpha_E$ coupling for a 10 MeV Elko at the 14 TeV LHC.
\begin{center}
	\includegraphics[scale=0.6]{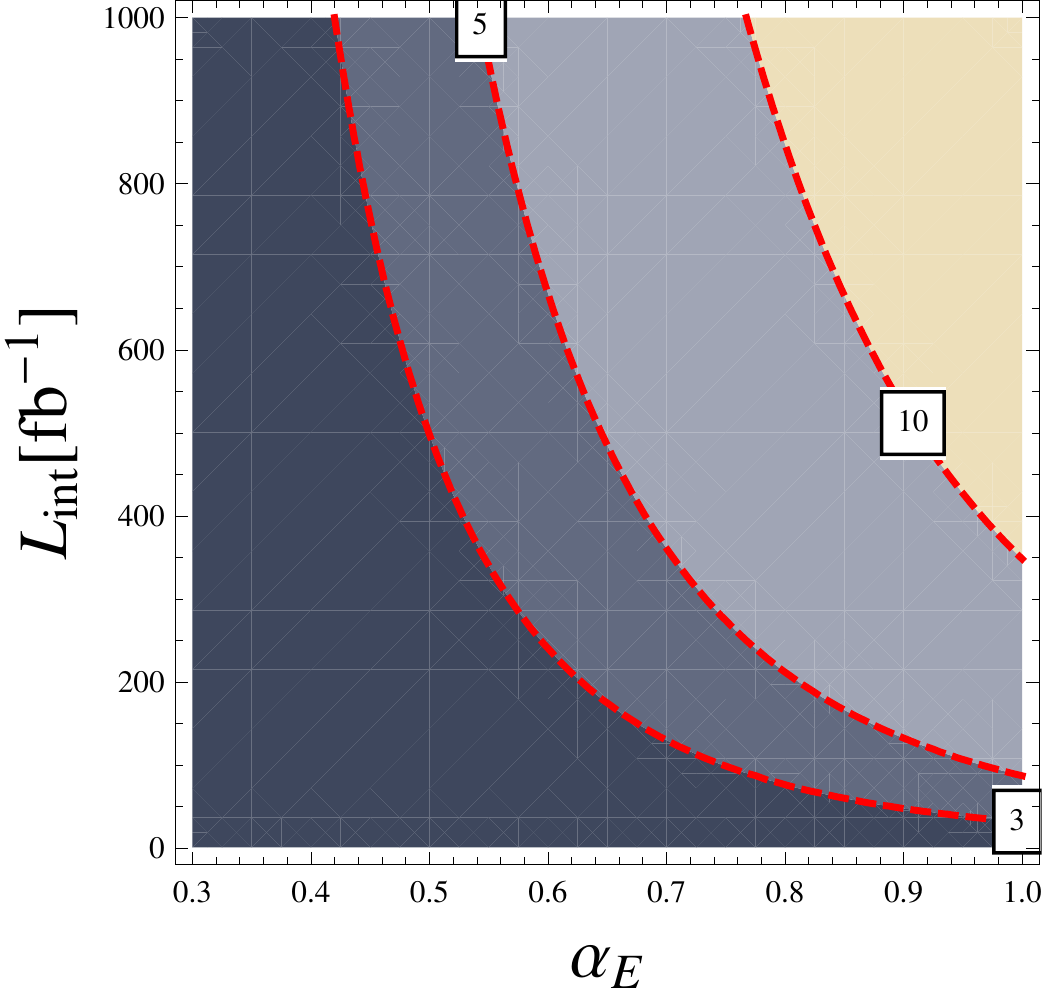}
	\captionof{figure}{ Significance as a function of $\alpha_E$ and the integrated luminosity in $fb^{-1}$.}
	\label{lumilam}
\end{center}

With 1 ab$^{-1}$, $HHE$ couplings as small as $\sim 0.4$ can lead to an evidence signal of $3\sigma$ at the LHC, and a $5\sigma$ discovery is possible for an enhanced coupling of $0.5$. If $\alpha_E \gtrsim 0.6$, an integrated luminosity of $\sim 500$ fb$^{-1}$ suffices for discovery of the Elko spinor at the 14 TeV LHC. 
Further detailed analysis are required, including a complete simulation of the detectors in order to confirm the precise limits on the coupling constant.

\section{Conclusions}

The Elko field, an spinor with mass dimension one, is a natural candidate for the constituent of dark matter. Its interactions with the Higgs boson open the possibility of discovery at colliders, as the LHC. 

We have investigated in this letter two scenarios for Elko interactions with Higgs bosons: the quartic coupling scenario, for which a 10 MeV Elko is shown to give rise to the right relic abundance as measured by WMAP, and the triple coupling scenario, where the Elko mass is generated trough the electroweak symmetry breaking mechanism.

The quartic scenario is very challenging even at the 14 TeV LHC for all the most promising channels. On the other hand, if triple couplings are present, the Elko can be easily discovered in the $pp\rightarrow \ell^+\ell^-+\eslash$ channel. For example, with a $HEE$ coupling of 0.5, a 10 MeV Elko discovery is possible after 1 ab$^{-1}$. However, couplings of order 0.6 or larger can be probed with up to 500 fb$^{-1}$.

In the triple coupling scenario, the Elko search can benefit from dark matter searches in mono-$Z,W$, monojet and monophoton channels at the 14 TeV LHC. Recent analyzes based upon the LHC7 and LHC8 is not likely to bound the Elko coupling however, since the signal cross sections are too small. 

Nevertheless, the 14 TeV LHC may open other possibilities as the WBF channel and an improvement on the bound on the Higgs invisible decay branching ratio.

\begin{acknowledgements}
We would like to thanks Profs. Alekha Nayak and Pankaj Jain for the careful reading of the manuscript and the important suggestions concerning the calculations. This research was supported by resources supplied by the Center for Scientific Computing (NCC/GridUNESP) of the 
S\~ao Paulo State University (UNESP). JMHS thanks to CNPq for partial financial support (482043/2011-3; 308623/2012-6).
\end{acknowledgements}


\begin{thebibliography}{99}
\bibitem{elko}
D.~V.~Ahluwalia and D.~Grumiller,
%''Spin half fermions with mass dimension one: Theory, phenomenology, and dark matter,''
JCAP {\bf 0507}, 012 (2005).

\bibitem{elko2}
D.~V.~Ahluwalia and D.~Grumiller,
%''Dark matter: A spin one half fermion with mass dimension one?,''
Phys.\ Rev.\ D {\bf 72}, 067701 (2005).

\bibitem{elksplb} D. V. Ahluwalia, C-Y. Lee, and D. Schritt, Phys. Lett. B {\bf 687}, 248 (2010).

\bibitem{ult} D. V. Ahluwalia,  arXiv:1305.7509 [hep-th] (2013).

\bibitem{Trefzger:1999bna} 
  T.~M.~Trefzger and K.~Jakobs,
  %``SM Higgs Searches for $H\rightarrow WW^{(*)}\rightarrow l^+\nu l^-\nu$ with a Mass between 150-190~GeV 
  at LHC,''
  ATL-PHYS-2000-015.

\bibitem{madgraph} 
  M.~Herquet and F.~Maltoni,
  %``MadGraph/MadEvent : A multipurpose event generator,''
  Nucl.\ Phys.\ Proc.\ Suppl.\  {\bf 179-180}, 211 (2008).
  %%CITATION = NUPHZ,179-180,211;%%

\bibitem{Conte:2012fm} 
  E.~Conte, B.~Fuks and G.~Serret,
  %``MadAnalysis 5, A User-Friendly Framework for Collider Phenomenology,''
  Comput. \ Phys. \ Commun. \ {\bf 184}, 222 (2013).
\bibitem{pythia} 
  T.~Sjostrand, S.~Mrenna and P.~Z.~Skands,
  %``A Brief Introduction to PYTHIA 8.1,''
  Comput.\ Phys.\ Commun.\  {\bf 178}, 852 (2008)
  [arXiv:0710.3820 [hep-ph]].
\bibitem{atlas} ``ATLAS: Detector and physics performance technical design report. Volume 1,''
  CERN-LHCC-99-14.
\bibitem{atlas2} ``ATLAS: Detector and physics performance technical design report. Volume 2,''
  CERN-LHCC-99-15.
\bibitem{Murayama:1992gi} 
  H.~Murayama, I.~Watanabe and K.~Hagiwara,
  %``HELAS: HELicity amplitude subroutines for Feynman diagram evaluations,''
  KEK-91-11.

\bibitem{monoj} The CMS Collaboration, CMS PAS EXO-12-048.

\bibitem{Djouadi:1997yw} 
  A.~Djouadi, J.~Kalinowski and M.~Spira,
  %``HDECAY: A Program for Higgs boson decays in the Standard Model and its supersymmetric extension,''
  Comput.\ Phys.\ Commun.\  {\bf 108}, 56 (1998).

\bibitem{invhiggs} G. Belanger, B. Dumont, U. Ellwanger, J. F. Gunion and S. Kraml, Phys. Rev. D {\bf 88}, 075008 (2013).

\end{thebibliography}
\end{document}